\documentclass[aps,plb,preprintnumbers,nopacs,nofootinbib,onecolumn]{revtex4}
\usepackage{graphicx}
\usepackage{bm}
\usepackage{amsmath}
\usepackage{amssymb}
\usepackage{amsthm}
\usepackage{color}
\usepackage[usenames,dvipsnames]{xcolor}
\usepackage{latexsym}
\usepackage{epsfig}
\newcommand{\lp}{\left(}
\newcommand{\rp}{\right)}
\newcommand{\lb}{\left[}
\newcommand{\rb}{\right]}

\newcommand{\der}{\hat{\nabla}}

\newcommand{\be}{\begin{equation}}
\newcommand{\ee}{\end{equation}}
\newcommand{\lsim}   {\mathrel{\mathop{\kern 0pt \rlap
  {\raise.2ex\hbox{$<$}}}
  \lower.9ex\hbox{\kern-.190em $\sim$}}}
\newcommand{\gsim}   {\mathrel{\mathop{\kern 0pt \rlap
  {\raise.2ex\hbox{$>$}}}
  \lower.9ex\hbox{\kern-.190em $\sim$}}}
\newcommand{\bw}{\begin{widetext}\begin{equation}}
\newcommand{\ew}{\end{equation}\end{widetext}}
\newcommand{\beq}{\begin{equation}}
\newcommand{\eeq}{\end{equation}}
\newcommand{\bea}{\begin{eqnarray}}
\newcommand{\eea}{\end{eqnarray}}

\newcommand{\e}{{\rm e}}
\newcommand{\f}{{\rm f}}
\newcommand{\s}{{\rm S}}

\newcommand{\poincare}{{Poincar\'e}}

\begin{document}

\title{Towards a gauge theory of frames}

\author{Tomi S. Koivisto}
\email{tomi.koivisto@nordita.org}
\affiliation{Nordita, KTH Royal Institute of Technology and Stockholm University, Roslagstullsbacken 23, 10691 Stockholm, Sweden}

\preprint{NORDITA-2015-125}
\date{\today}

\begin{abstract}

The equivalence principle postulates a frame. This implies globally special and locally general relativity. 
It is proposed here that spacetime emerges from the gauge potential of translations, whilst the Lorenz symmetry is gauged into the interactions of the particle sector. This is,
though more intuitive, the opposite to the standard formulation of gravity, and seems to lead to conceptual and technical improvements of the theory. 

\end{abstract}


\maketitle

\section{Introduction}

From particle physics, we have learned that local interactions are well described by the Yang-Mills theory in \poincare-invariant spacetime \cite{'tHooft:1995gh}. 
A theory, in general, can be formulated as a statement $\s$, so that classical physics follow from the invariance of $\s$, and quantum physics incorporate the effect of all possible deviations according to their probability. Unfortunately, it is still conventional to formulate $\s$ as a spacetime integral $\hat{\mathcal{I}}$ over a function $\mathcal{L}(\phi)$ of fields $\phi$,
\be \label{action1}
\s = \hat{\mathcal{I}} \mathcal{L}(\phi)\,,
\ee
even though there is evidence that spacetime is not fundamental. Gravity \cite{Misner:1974qy} is incorporated into this effective description by curving the spacetime $x^\mu$ with a metric $g_{\mu\nu}(x)$, and the integral 
operator $\hat{\mathcal{I}}$ is then written as
\be \label{integralm}
\hat{\mathcal{I}}=\int d^{D}x \sqrt{-g}\,.
\ee
When this description is breaking down at high energy or small distance scales, there are various hints as to how it should be modified. We can consider what effectively happens to each of the symbols in the right hand side of the integral (\ref{integralm}) near the limits of its validity: the gravitational interaction $g$ will become nonlocal \cite{Biswas:2011ar,Modesto:2011kw}; instead of the $-$, the opposite sign could be at least more convenient for the signature \cite{Gibbons:1978ac,Perez:2000bf} that can be also dynamical \cite{0264-9381-27-4-045007,Carlini:1993up}; the integral $\int$ might be better described by a discrete sum \cite{Rovelli:1997yv,Ambjorn:2005qt}, the coordinates $x$ by noncommutative variables \cite{Seiberg:1999vs,Szabo:2006wx}, the $d$ by some topologically nontrivial \cite{AmelinoCamelia:2008qg,Hossenfelder:2014hha} and the $\sqrt{}$ by some fractal generalisation \cite{Benedetti:2008gu,Calcagni:2011kn} since, finally, the number $D$, that is $D=4$ at larger distances, will become $D=2$ \cite{Ambjorn:2005db,Modesto:2009qc}. 

Physics in spacetimes and mathematics on differentiable manifolds cannot thus be straightforwardly applied to elaborate the gravitational theory. 
Revisiting it from the perspective of gauge theory turns out to be illuminating \cite{Blagojevic:2002du,Aldrovandi:2013wha}. Hermann Weyl introduced the vielbein as a more fundamental object than the metric, a mapping from the field space to the external space. Perhaps, we needed not to assume any spacetime to begin with, but could rather consider the $\hat{\mathcal{I}}$ as a collector of elements on some abstract structure which has a symmetry. 
Spacetime would then emerge as the gauge potential of this symmetry, the vielbein, or {\it a frame}.  
From the gauge perspective, another well-known but much less well understood shortcoming of the effective metric starting point (\ref{integralm}) becomes also evident, to wit that it misses half of the picture by neglecting the dual of the curvature, {\it the torsion}. One realises that in the conventional formulation of gravity theory, the gauge potential is that of the internal Lorenz spin action (field strength of which is identified as the Riemannian curvature), 
whereas, intuitively, spacetime would be generated by translations (field strength of which is geometrically the torsion). In the exceptional case of the Einstein-Hilbert action though, there is a duality between these formulations.  
We believe this symmetry is not accidental, but also, that the geometric interpretation of gravity as a curvature of spacetime it admits may have been misleading us more than expected. 

Besides the proper gauge, the proper fields of the gravitational interaction may not have been identified. The former would be a misleading choice, the latter a mistake. 
We have to point attention to a subtle defect of the Poincar\'e gauge theory of gravity 
(PGT) \cite{Hehl:1976kj}: the action that reduces to the Einstein-Hilbert in the metric gauge, is actually inhomogeneous. 
The more fundamental repercussions of this may have remained in hiding since, on one hand, as the non-covariant term appears in the action as a total derivative, it can be ignored in the teleparallel equivalent of general relativity (TEGR) \cite{Hayashi:1979qx}, and on the other hand, as it is torsional, it does not play a dynamical role in general relativity (GR).
The term fails covariance because the field strength of the Lorenz spin has to be projected to the fully covariant space with vielbeins (which as gauge potentials transform non-covariantly). This is not a failure of the gauge principle, but rather a hint towards the correct theory. When completing the work of Utiyama \cite{Utiyama:1956sy}, Kibble  \cite{Kibble:1961ba} gauged only a piece of the Lorenz structure separately with the inhomogeneous part of the group, and as commented earlier, it is the latter whose gauge field de facto couples to the spin structure of matter in PGT. 

Both aesthetic and technical reasons compel us to consider the full action of the first semi-direct factor in $SO(3,1) \rtimes T(3,1)$, and use the second to generate space. 
The theory that emerges is a {\it torsion gauge theory of spacetime frames with pseudoteleparallelly unified matter sector}. 
Thus, the $[SU(2)\otimes SU(2)]/\mathbb{Z}_2$ is associated to the particle sector, and indeed it is rightaway tantalisingly close to what we would seem to precisely require there. Such a unification has been pursued, since the insights that led to the first gravity theory (with electromagnetism as an extra dimension), the first gauge theory (electromagnetism as the conformal symmetry) and the introduction of teleparallelism (electromagnetism as extra entries in the vierbeine), from the geometric standpoint of what we call the "metric gauge". The first teleparallel unification was unsuccesfull as it obtained the Maxwell theory as a coordinate effect that could be transformed away. In PGT, that is a theory of gravitation alone, such transformations imply tidal forces that lead to the conception that spacetime, in addition to being curved by matter, has to ''twirl'' as well. This new gravitational force appears superfluous in our scheme, wherein the Lorenz force is placed in situ, and gravity is given by the "twirling". In hindsight, this picture has been suggested by such findings as the gravitoelectromagnetic analogy \cite{Mashhoon:2003ax}, the resolution of the gravitational energy density in TEGR \cite{deAndrade:2000kr} and its problem with spin coupling \cite{Obukhov:2002tm}. 

Here, we still write the theory in the traditional, what could be called the "metric gauge" (or rather a pseudogauge, since it is not fully covariant in standard theories) formulation in the following Section \ref{theory}, introducing then multiple sectors of spacetimes and fields. We can then review and generalise extended gravity theories in the metric gauge from the new perspective in Section \ref{applications}, where we are also lead to define a generalised concept of spacetime. The conclusion $\s$ of Section \ref{conclusion} is the gauge theory of frames we have already specified above. 
  
\section{Theory}
\label{theory}

The operator $\hat{\mathcal{I}}$ sums over the totality of elements on the absolute Frame, and is thus independent of any ordering.
If we are less pretentious and take $\hat{\mathcal{I}}$ as an integral $\int dx$ over a Minkowski space $x$, we have then diffeomorphism invariance. In general, the transformation $\delta_\chi$ generated by the operator $\chi$ leaves $\s$ unchanged if
\be \label{diffeo}
\delta_\chi \mathcal{L} = \frac{\partial\mathcal{L}}{\partial \phi} \delta_0 \phi + \chi \cdot \partial \mathcal{L} + \lp\partial\cdot\chi\rp  \mathcal{L} = 0\,.
\ee
where $\partial$ could be interpreted as (minus) the translation generator, and $\delta_0$ is the action of $\chi$ to the field. 

In the case of a Poincare transformation $\delta_P$, the field action is 
\be \label{poincare}
\delta_P \phi = \lp -\xi \cdot \partial + \frac{1}{2}\omega \cdot \Sigma\rp\phi\,,
\ee
where $\Sigma$ does the internal spin action with parameterised by $\omega$ and, following Kibble \cite{Kibble:1961ba}, we have absorbed the translations and rotations into $\xi$, such that  
$\xi = (\epsilon + \frac{1}{2}\omega \cdot x)$, where $\epsilon$ is a pure translation. Note that we can always absorb $\omega$, but then cannot always treat $\xi$ as completely independent.

Now we have that
\be \label{failure}
\lb \delta_P, \partial \rb \phi = - \lp \partial \xi \rp\cdot\phi + \frac{1}{2}\partial \lp \omega \cdot \Sigma\rp \phi\,,
\ee
but can define a covariant derivative $\nabla = \partial - A$, such that $\lb \delta_P, \nabla \rb = 0$, by introducing the compensating field $A$ which transforms as
\be
\delta_P A = \lb A, \partial \rb - \lp\partial\cdot\xi\rp\cdot A - \frac{1}{2}\partial \lp \omega \cdot \Sigma\rp\,. 
\ee
We have then gauged $\Sigma$ into the action (\ref{action1}) as $\s \rightarrow \hat{\mathcal{I}} \mathcal{L}(\phi, \nabla)$. One can make a note now that the field strenght will be interpreted as the Riemann curvature tensor
\be \label{Riemann}
R = [\nabla,\nabla]\,,
\ee
in the conventional metric formulation (\ref{integralm}).
 
Let us next consider a series $\xi^{\Upsilon}$ of transformations. We can then define the corresponding set of objects $\nabla^\Upsilon$ with gauge fields $A^\Upsilon$,
\be 
\s = \mathcal{I}_\Upsilon \mathcal{L}^\Upsilon(\phi, \nabla) \,,
\ee
each performing covariant operations upon the fields of their respective sectors such that we read $\mathcal{L}^\Upsilon(\phi,\nabla)=  \mathcal{L}^\Upsilon(\phi^\Upsilon, \nabla^\Upsilon)$.
Thus, we can also associate an index structure to $\phi$. 

To further gauge the translation symmetry, we note from equation (\ref{failure}) that the failure of $\nabla^\Upsilon_\mu$ to transform \poincare  -invariantly is linear. It can be therefore eliminated by introducing the covariant operators $\der$ and the set of compensating fields $\f_{\Upsilon}$ such that
\be
\der\phi = \f_{\Upsilon} \nabla^{\Upsilon}\phi\,,  \quad \lb\der, \delta_0\rb\phi=0\,,
\ee 
which fixes the transformation law for the mapping $\f_{\Upsilon}$:
\be
\lb \delta_P,  \f_{\Upsilon} \rb = \frac{1}{2}\omega\cdot \Sigma f_{\Upsilon} + f_{\Upsilon}\cdot \partial\xi - \xi\cdot\partial f_{\Upsilon}\,.
\ee
The complications arising from the absorption of $\omega\cdot x$ into $\xi$ in the starting point (\ref{poincare}) could be anticipated from the nonlinear coupling of the canonical potentials.

We will further consider a series $\omega^{A}$ of transformations and their commuting operators $\nabla^A$. We have then a set of fields $\f_{\Upsilon}^A$, and can write a multiple-gauged theory 
\be \label{action2}
\s = \hat{\mathcal{I}}_{\Upsilon}^A \mathcal{L}^{\Upsilon}_A(\phi, \der) = \hat{\mathcal{I}}_{\Upsilon} \det\lb \f^{(-1)}\rb^A \mathcal{L}^{\Upsilon}_A(\phi, \der)\,,
\ee
where the determinant-like operator needs to be included to compensate for the $\partial\cdot\xi$-term in equation (\ref{failure}). 
The field strengths for the gauge fields can be defined via the projections of the $\nabla$-commutator
\be
\hat{F}^{AB}   \equiv  \f^A_{\Upsilon} \f^B_{\Psi} \lb \nabla^\Upsilon, \nabla^\Psi \rb\Sigma^{-1}\,, \label{strength} \quad
\hat{T}_A^{\phantom{A}BC}   \equiv  \f^B_\Upsilon \f^C_\Psi \lb \nabla^\Upsilon, \f^{(-1)\Psi}_A\rb\,, \quad \text{where} \quad A=B=C\,.
\ee
We can then show that  $[\hat{\nabla}, \hat{\nabla}] = \hat{F}\cdot\hat{\Sigma}-\hat{T}\cdot\der$. The unorthodox choice of generators manifests finally here in that the covariant object does not combine into a quadratic Lagrangian, whilst the projected object is not fully covariant. 

The action is then completed by taking into account the interactions in the different gauge sectors,
\be
\s = \hat{\mathcal{I}}_{\Upsilon} \det\lb \f^{(-1)}\rb^A  \mathcal{L}^{\Upsilon}_A(\hat{F},\hat{T},\phi, \der)\,,
\ee
and we can discuss the physical interpretations of the theory.

\section{Models}
\label{applications}

In this section, we revisit and generalise theories in the ''metric gauge'' from the new perspective. 

\subsubsection{$0\times\partial, 0\times\Sigma$:  (SR)}

We proceed by systematic repetitions of the gauging maneuver and present one example of a model from each class of theories. 

\subsection{Single sector}
\label{one}

Since $\xi$ can absorb $\omega\cdot x$ in (\ref{poincare}), the (restricted) rotations can simulate gravity (of the Frame theory). In TEGR literature the viewpoint has been captured as an electromagnetic analogy of GR. The latter is actually the theory of the spin $\Sigma$, that would naturally have more (and not the most suitable) degrees of freedom (6) than needed (4). This is  why the Einstein-Hilbert is second order and ultra-sensitive to modifications. One could also formulate an equivalent of GR by gauging only the rotations $\omega$.  
 
\subsubsection{$1\times\Sigma$: $\mathcal{L} = R$ (GR)}

The problems of GR are be understood when it is recognised as a teleperpendicularly misinterpreted Yang-Mills theory of the (incomplete) Lorenz rotation.

\subsubsection{$1\times\partial$: $\mathcal{L} = T$ (TEGR)}

The three quadratic invariants one can compose from $\hat{T}$ in equation (\ref{strength}) are the axial $a^2$, polar $v^2$, and tensorial $t^2$ torsions \cite{Hayashi:1979wj}, whose linear combination $T=3a^2/2-2(v^2-t^2)/3$ appears in the action of TEGR. This $T$, like any combination $f(a^2,v^2,t^2)$ is gauge-invariant. Curvature plays no role in those theories (for gravity). The Weitzenb\"ock condition would appear only as an unphysical restriction that actually adds curvature into the theory - in the form of a Lagrange multiplier that spoils the covariance (unless manipulated correctly as a local inertial frame). This had not been very clear in the current literatures.

As emphasised, the teleparallel paradigm has had it right all the time \cite{Aldrovandi:2013wha}. A question that has cristallised in the Frame theory is the role of the two other torsional modes besides the graviton: only a part of $T(4)$ is needed for an effective geometry.

\subsubsection{$1\times\Sigma + 1\times\partial$: $\mathcal{L} = \hat{F} = R - T + 2\nabla v = F_\Sigma + F_\xi$ (PGT)}

The last term in the action of PGT is a total derivative of the polar torsion defined as the trace of $\hat{T}$ in the equation (\ref{strength}), and above we have introduced a suggestive notation.

\subsection{Two sectors}
\label{two}

The multiple-gauged theories can accommodate many distinct but possibly overlapping spacetimes, and many distinct but possibly interacting gauge fields; the effective spacetime can then be seen as a combination of several overlapping geometries. Interesting examples of such theories are the massive \cite{deRham:2010kj,deRham:2014zqa} and the bimetric \cite{Hassan:2011tf,Hassan:2011zd} theories of gravity, that introduced the three ghost-free couplings between two metrics. Consider the determinant element $\lb \det{\e}\rb^A$ in the theory (\ref{action2}) with doubled gravity. If $D=4$, the determinant is a product of 4 tetrads, and each can be chosen from either of the sectors. We have then in fact the total of $16$ distinct ''spacetimes'',
$\lb \det{\e}\rb^A$, and know that at least the simplest $\mathcal{L}_A$, a cosmological constant $\Lambda_A$, can be safely added into each of them. In the metric picture, only combinations and not permutations can be coupled differently, and we have five relevant combinations of the cosmological constants. The cross-terms are of particular interest,
\be
V_{(1,2)} = \epsilon_{a b c d}\lb 
\Lambda_1 \e_{(2)}^a\wedge \e_{(1)}^b\ \wedge \e_{(1)}^c\ \wedge \e_{(1)}^d +
\Lambda_2 \e_{(2)}^a\wedge \e_{(2)}^b\ \wedge \e_{(1)}^c\ \wedge \e_{(1)}^d +
\Lambda_3 \e_{(2)}^a\wedge \e_{(2)}^b\ \wedge \e_{(2)}^c\ \wedge \e_{(1)}^d
\rb\,,
\ee
since these are the three known ghost-free interactions. Let us now contemplate only this (bi)metric picture without torsion. From the Frame we see in a metric, a part of the hadronic structure \cite{Isham:1971gm}.

\subsubsection{$2 \times \Sigma$: $\mathcal{L}=  R_{(1)} + V_{(1,2)}$ (MG)}

The action of massive gravity equips only one the metrics with dynamics, and the other remains a trivially invariant ''reference metric''. 

\subsubsection{$1 \times \partial + 2 \times \Sigma$: $\mathcal{L}=  R_{(1)} + R_{(2)} + V_{(1,2)} + \mathcal{L}^{(1)}_{(1)} + \alpha\mathcal{L}^{(1)}_{(2)}$ (2G)}

The bimetric massive gravity when $\alpha = 0$ is GR with a consistently coupled extra spin-2 field.
When $\alpha \neq 0$, we were lead to reconsider the nature of spacetime, then in the context of Finsler geometry \cite{KOIVISTO:2013jwa,Akrami:2014lja}. (However, the symmetric matter couplings are not ghost-free \cite{Yamashita:2014fga,Noller:2014sta}, but possibly viable composite coupling to matter, $\mathcal{L}_{(1,2)}$, could be constructed \cite{Heisenberg:2015iqa,Heisenberg:2015wja}.) 

\subsubsection{$2 \times \partial + 2 \times \Sigma$: $\mathcal{L}=  R_{(1)} + R_{(2)} + V_{(1,2)} + \mathcal{L}^{(1)}_{(1)} + \mathcal{L}^{(2)}_{(2)}$ (d2G)}

Twin matters can be coupled consistently, if to separate sectors \cite{Aoki:2013joa,Lagos:2015sya}. 

\subsection{Different families of sectors}

Besides multiple generations, we can consider different gauging prescriptions, say, the $\omega$-generated equivalent of GR, an orthochronous variation of the transformation or the refined torsion gravity with the tensorial (usual graviton) and the polar and axial coupled to distinct sectors. 
The {\it gauge-gravity equivalence} established by the Frame theory invites to study its implications. Relativity is particularised, and particles relativised: the particle theory is different in each background.
For example we could predict its workings in the (a)dS space, which we know it could be mapped into the theory in a flat spacetime with one dimension less. VR shows that rather than SR, VSR resembles the global symmetry of our standard model \cite{Cohen:2006ky}. That is now connected to (a)symmetries in cosmology, which could be interesting in principle \cite{Brans:1961sx} and observationally \cite{Ade:2015hxq}. 

\section{Conclusion}
\label{conclusion}

The Frame theory, or {\it the unified gauge theory of frames} is a radically new paradigm that eliminates the spacetime by unifying it intra the other interactions.
It used to be that ''spacetime tells matter how to move; matter tells spacetime how to curve'', we propose that ''spacetime is lost in translations; matters
depend on the angle''.

The gauge theory of translations is the (thermo)dynamics of spacetime, equivalent to the energy and momentum of matter. In GR the Riemann curvature $R$ plays the role of gravity, but in the Frame theory the $R$ is seen as (a gauge-noninvariant version of) the Lorenz forces responsible for the strong interaction. Teleparallelism has to be thus accepted, not a priori with any Weitzenb\"ockian restriction, and not as GR in a different gauge in Einstein-Cartan spacetime as in PGT, but as the (more) physical interpretation of the (pure) gravity sector. To our knowledge, it has not been much investigated whether the TEGR action or its gauge-invariant torsional generalisations would be more amenable to quantisation than the Einstein-Hilbert that is second order derivative. It remains to be seen also, whether a subtle revision of the ''gravity''-matter interaction would render it insensitive to a constant background source, in line with the other Yang-Mills theories of fundamental interactions, or what is the interpretation of a $\Lambda$ in Frame theory.    

\bibliography{refs}

\end{document}